\documentclass[
 reprint,
 preprintnumbers,
 amsmath,
 amssymb,
 aps,
 prstab,
 floatfix,
]{revtex4-1}

\usepackage{graphicx}			
\usepackage{verbatim}			
\usepackage{wasysym}			
\usepackage{mathtools}			
\usepackage[makeroom]{cancel}		
\usepackage{dcolumn}			
\usepackage{hyperref}			
\usepackage{color}			

\begin{document}

\title{Beam Breakup Mitigation by Ion Mobility in Plasma Acceleration}
\author{A. Burov, S. Nagaitsev and V. Lebedev}
\affiliation{Fermilab, PO Box 500, Batavia, IL 60510-5011}
\date{\today}

\begin{abstract}
Moderate ion mobility provides a source of damping in the plasma wakefield acceleration, which may serve as an effective remedy against the transverse instability of the trailing bunch. Ion mobility in the fields of the driving and trailing bunches is taken into account; the related effects are estimated for the FACET-II parameters.
\end{abstract}

\maketitle


Plasma wakefield acceleration (PWA) suggests extremely high acceleration fields, so it is no surprise that this area of research attracts interest of groups working on future linear  colliders, giving rise to many publications, targeted at resolution of multiple interrelated problems in this challenging area. 
A special subset of these problems is associated with stability of both driving (accelerating) and trailing (accelerated) bunches. The latter problem appears to be harder than the former, since mismatches at every change of the driving bunch between positions of the two bunches produce initial kicks for the instability development along the acceleration line for one and the same trailing bunch. From a very general point of view, the PWA trailing bunch instability belongs to the family of similar effects in linacs. Due to interactions with the surroundings, dipole perturbations at the head of the bunch leave electromagnetic wake fields behind, thus acting on the bunch tail. The kick felt by the test particle from a unit dipole moment of the leading particle is known as the wake function, $W_\perp(\xi)$, where $\xi$ is the separation between the particles, see e.g. Ref.~\cite{chao1993physics}. As a result of this head-to-tail interaction, the tail dipole oscillations may grow more and more, leading to the emittance degradation. This sort of {\it unbounded convective instability}~\cite{Burov:2018pjl} is known as the {\it beam breakup}. Here we are considering the acceleration of a short electron bunch in the blowout regime, a regime in which the fields of the driver (laser or an electron bunch) are so intense that they expel all plasma electrons, creating a cavity filled with ions only~\cite{PhysRevA.44.R6189}. The longitudinal and transverse electric fields inside this cavity are used to accelerate and focus the trailing electron bunch. The transverse wake fields are very sensitive to the aperture radius, which is the plasma bubble radius at the bunch location, $r_b$, for the PWA case: for the short bunches of the interest, the wake function is inversely proportional to the fourth power of this radius, $W_\perp(\xi) \simeq 8 \xi \Theta (\xi)/r_b^4$, where $\Theta(\xi)$ is the Heaviside theta-function; details on that can be found e.g. in Refs.~\cite{PhysRevAccelBeams.21.041301, PhysRevAccelBeams.21.071301}. To get the desired high acceleration, the plasma bubble has to be small, typically $r_b \simeq 50-100\mu$m, compared with $1-2$cm for conventional colliders; thus, with the fourth power of the aperture in the transverse wake, the transverse instability is by necessity one of the main obstacles for the PWA colliders. From this, one may correctly conclude that there must be a conflict relation between energy efficiency and beam stability for the PWA: while the former requires smaller bubbles, the latter is lost with them. Such {\it efficiency-instability relation} has been recently formulated and proved in Ref.~\cite{PhysRevAccelBeams.20.121301}; here we reproduce the result for the reader's convenience. 

Let $\eta_P < 1$ be the PWA energy efficiency, i.e. the ratio of the power of the trailing bunch acceleration to the power of the driving bunch deceleration. Further, let the wake parameter $\eta_t$ be the ratio of particle defocussing by the transverse wake fields to the main focusing of the bunch electrons by the ions inside the bubble for the case when the particle is located at the bunch end and all particles are transversely  displaced by the same distance. The efficiency-instability relation of Ref.~\cite{PhysRevAccelBeams.20.121301} states that  
\begin{equation}
\eta_t \approx \frac{\eta_P^2}{4(1-\eta_P)}\,.
\label{EffInstabRel}
\end{equation}  
By virtue of this relation, an increase in the efficiency inevitably entails the corresponding elevation of the wake relative strength, thus bringing the bunch closer to the instability threshold. An introduction of sufficiently large transverse frequency spread suppresses this instability. A particular case of such stabilization is BNS damping~\cite{Balakin:1983sc}, where the transverse frequency varies along the bunch by means of the energy modulation. For the PWA though, there is a natural mechanism of the frequency spread, associated with the ion mobility in the Coulomb fields of the driving and trailing bunches. This mobility causes variation of the ion density $\delta n_i$, providing a spread of the transverse frequencies. 

Ion mobility in the Coulomb field of the driving bunch causes a nonlinear focusing of the trailing bunch, enhanced proportionally to the distance between the bunches. Assuming small ion density perturbation and the distance $l$ to be much larger than the bunches' lengths, the nonlinear detuning of the trailing electron with the rms offset $\sigma_t$ can be calculated as
\begin{equation}
\frac{\delta \omega_d}{\omega_\perp} = \frac{3N_d Z_i r_p \sigma_t^2 l}{16 A_i \sigma_d^4}\,; \quad \sigma_t \leq \sigma_d\,. 
\label{domegad}
\end{equation}  
Here $N_d$ is the number of electrons in the driver, $Z_i$ and $A_i$ are the ion's charge and mass numbers, $r_p$ is the proton classical radius, $\sigma_d$ is the rms transverse size of the driving bunch. This estimation assumes that ions are far from collapse, $ N_d Z_i r_p l /A_i  \ll \sigma_d^2 $, see Ref.~\cite{PhysRevLett.95.195002}. The coherent amplitude $\bar{x}$ of a Gaussian ensemble of oscillators with such a frequency spread decreases with time $t\,$ so that $|\bar{x}(t)|=|\bar{x}(0)|/\sqrt{1+\delta \omega_d^2 t^2}$. After the oscillation period $T_\perp=2\pi/\omega_\perp$, the amplitude drops as 
$|\bar{x}(T_\perp)|/|\bar{x}(0)| = 1/\sqrt{1+(2\pi \delta \omega_d/\omega_\perp )^2}$.

Transverse stabilization of that sort was observed in simulations of Ref.~\cite{An:2017FACET2} with  $N_d=1.0\cdot 10^{10}$, rms transverse radii $\sigma_d=\sigma_t=0.52\,\mu$m, and the inter-bunch distance $l=150\,\mu$m. Substitution of these numbers yields the amplitude drop $|\bar{x}(T_\perp)|/|\bar{x}(0)|=0.60$, with an amazing agreement with the actual value.

The same ion mobility in the field of the trailing bunch leads to the ion density modulation along the bunch, which causes the corresponding frequency modulation $\delta \omega_t$, so that $\delta \omega_t/\delta \omega_d \sim (N_t/N_d)\, (\sigma_{zt}/l)\, (\sigma_d^4/\sigma_{t}^4)$. With acceleration, the transverse size of the trailing bunch shrinks, $\sigma_t^4 \propto 1/\gamma$; thus, at sufficiently high energy this size may become so small that the ion mobility in the field of the trailing bunch will be more important than such in the field of the driver. Let us assume now that this is the case, and estimate the stabilizing effect. For simplicity, we consider a constant-density bunch, which radius $b=2\sigma_t$ and full length $L_t=4\sigma_{zt}$. This yields the frequency modulation   
\begin{equation}
\frac{\delta \omega_t}{ \omega_\perp}= \frac{N_t Z_i r_p L_t \zeta^2}{A_i b^2} \,, 
\label{domit}
\end{equation}  
with $\zeta \equiv \xi/L_t$ as the dimensionless distance from the bunch head, $0 \leq \zeta \leq 1$.  The equation of motion for the bunch local offsets $X(\zeta,\,\mu)$ can be presented as~\cite{PhysRevAccelBeams.20.121301} 
\begin{equation}
\frac{\partial^2 X}{\partial \mu^2} + \left(1 + 2\frac{\delta \omega_t}{\omega_\perp}\right) X = 2\eta_t \int_0^\zeta X(\zeta')(\zeta - \zeta') d\zeta' \,.
\label{Xximu}
\end{equation}  
Here $\mu$ is time as the phase advance, $d\mu = k_p ds/\sqrt{2 \gamma}$, with $ds$ as the differential length along the bunch trajectory, $k_p=\sqrt{4\pi n_0 r_e}$ as the relativistic Debye length; $n_0$ is the plasma density and $r_e=e^2/(mc^2)$ is the electron classical radius.




The equation of motion~(\ref{Xximu}) can be further simplified with the slow amplitudes $x=X\exp{(i \mu)}$ and slow time $\tau = \mu \eta_t$. Then, it is transformed to the following equation on the slow amplitudes $x(\zeta,\tau)$, with the constant initial condition,
\begin{equation}
\begin{split}
& \frac{\partial x}{\partial \tau} = -i \kappa \zeta^2 x + 2 i  \int_0^\zeta x(\zeta-\zeta')\zeta' d\zeta' \,; \\
& x(\zeta,\,0)=1\,;  \quad 0 \leq \zeta \leq 1\,.
\end{split}
\label{xtauzeta}
\end{equation}  
Here we introduced the {\it modulation parameter}
\begin{equation}
\kappa \equiv \frac{2N_t r_i L_t}{b^2 \eta_t}=\frac{\mu_i^2}{\eta_t}= 1\,, 
\label{BNSi}
\end{equation}  
with $\mu_i = \sqrt{2N_t r_i L_t/b^2} \ll 1$ as the full phase advance of ion's oscillations in the field of the trailing bunch.

It follows from Eq.~\ref{xtauzeta} that if $\kappa=1$, the bunch initial deflection $x=1$ would result in its constant-amplitude oscillations as a whole, i.e. the slow amplitude would remain the same, $x(\zeta,\tau)=1$ in that case. In conventional linear accelerators, this condition is well-known as the BNS damping~\cite{Balakin:1983sc}, with the frequency modulation provided by the proper energy chirp. As it will be shown below, every frequency modulation makes the bunch more stable; the BNS condition $\kappa =1$ is sufficient, but not necessary for that. Due to  acceleration, the bunch shrinks, $b \propto \gamma^{-1/4}$, so the ion-related BNS parameter grows, $\kappa \propto \sqrt{\gamma}$. However, general properties of Eq.~(\ref{xtauzeta}) can be demonstrated neglecting this energy-dependence; so $\kappa = \mathrm{const}$ is assumed below.     
\begin{figure}[h!]
\includegraphics[width=\linewidth]{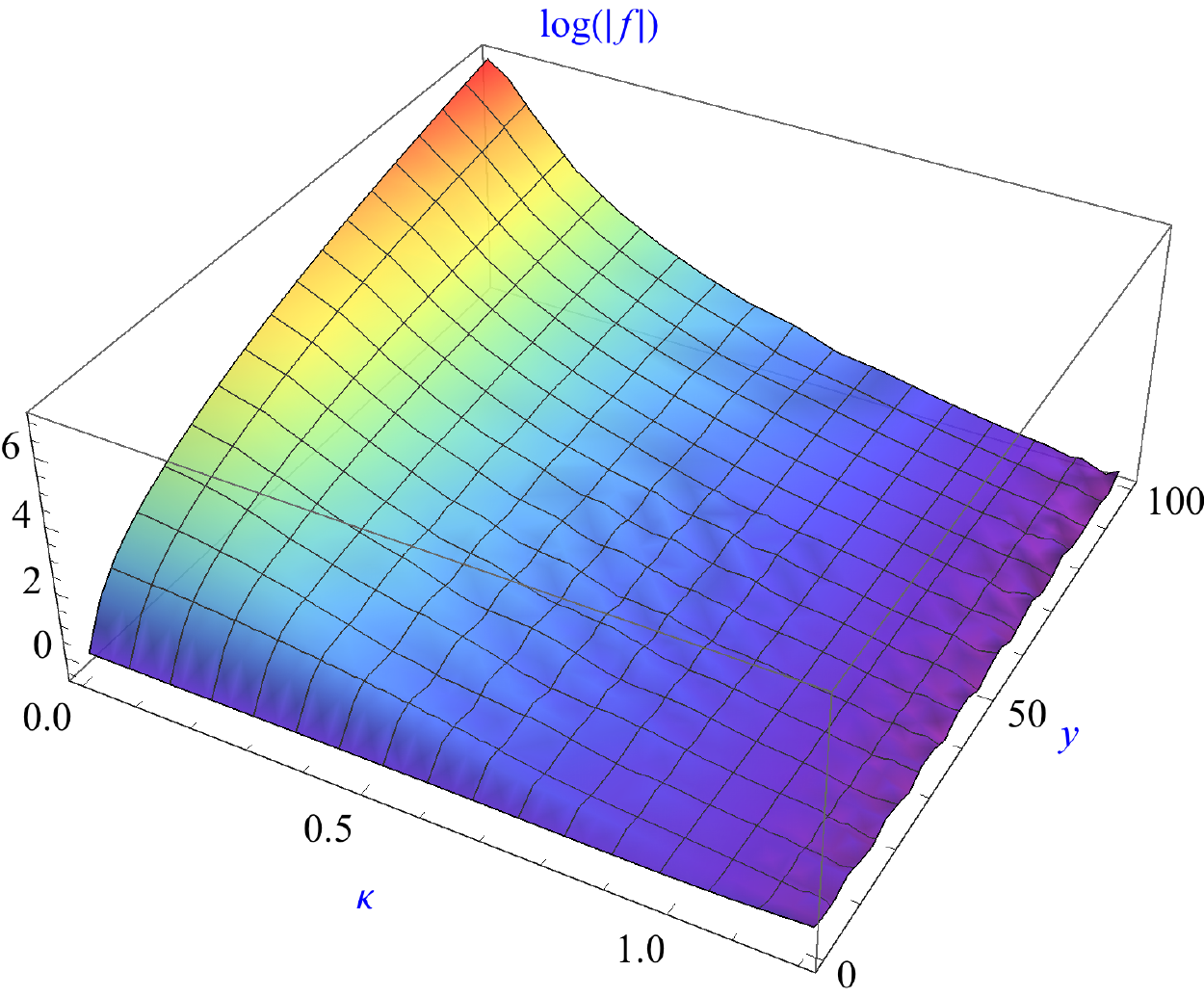}
\caption{\label{Fig:XKappaTau}
	Natural logarithm of the bunch oscillation amplitude $f(y) \equiv x(1,y)$, versus the scaled time $y \equiv \zeta^2 \tau$ and the frequency modulation parameter $\kappa$.   
	}
\end{figure}
With this simplification, Eq.~(\ref{xtauzeta}) does not depend on time directly, and it is straightforward to show that its solution has a scaling invariance: it depends on the space and time arguments $\zeta$ and $\tau$ as $x(\zeta,\tau)=x(1,\zeta^2 \tau) \equiv f(\zeta^2 \tau)$. In other words, the complex amplitude of the oscillations at position $\zeta$ and time $\tau$ is the same as at the bunch tail and earlier time $\zeta^2 \tau$. This means that the partial integro-differential equation~(\ref{xtauzeta}) with the specified initial condition is equivalent to an ordinary one. This ordinary integro-differential equation can be conveniently presented with the space-time argument $u \equiv \zeta \sqrt{2\tau}$. With  $x(\zeta,\tau)=g(u)$, Eq.~(\ref{xtauzeta}) reduces then to the following form
\begin{equation}
\frac{d\,g}{d\,u} = -i \kappa u g + \frac{2i}{u}  \int_0^u g(u-u') u'du'\,; \quad g(0)=1\,,
\label{gu}
\end{equation}  
At $\kappa =0$, i.e. without any damping, the solution at large argument, $u \gg 1$, asymptotically tends to $g(u) \simeq \exp(3\,i^{1/3}(u/2)^{2/3})$, omitting the pre-exponential factor.
\begin{figure}[h]
\includegraphics[width=\linewidth]{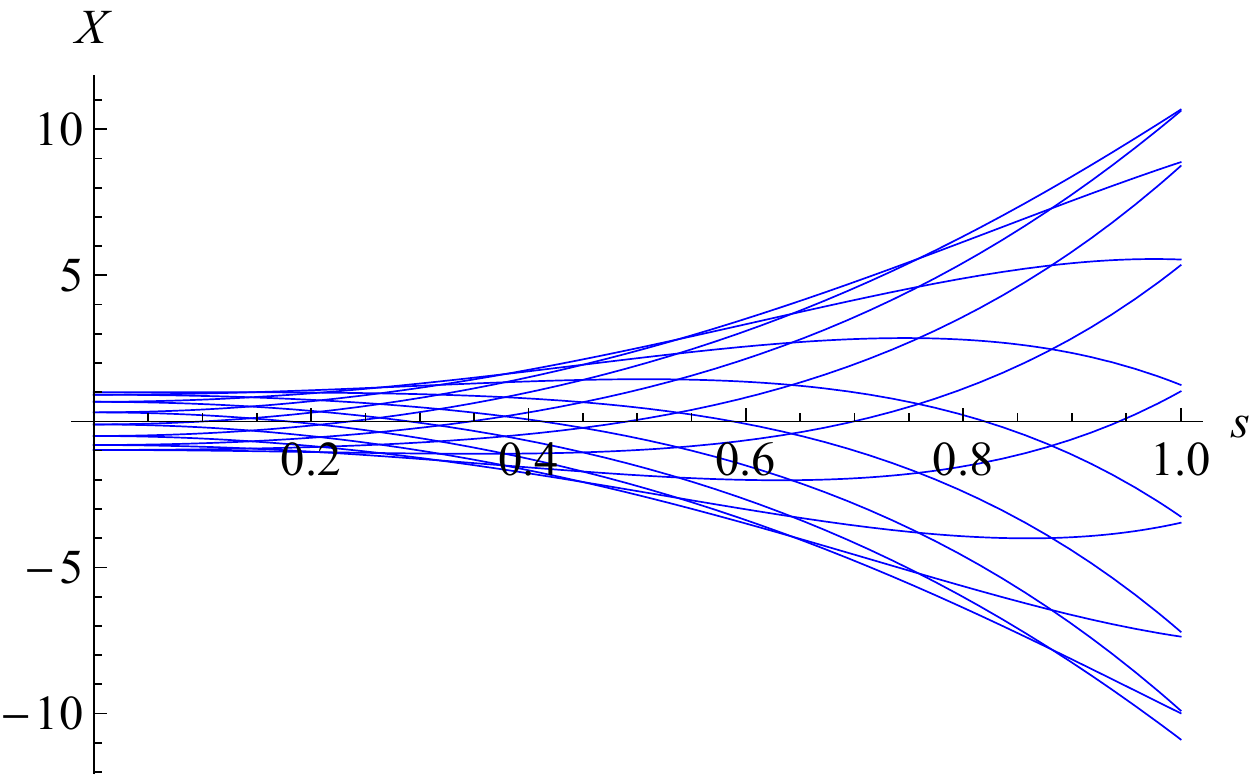}
\caption{\label{XKappa0Tau10}
	Stroboscopic image of the bunch oscillations $X(\zeta)$ for immobile ions, $\kappa=0$, at time $\tau=10$. 
	}
\end{figure}
\begin{figure}[h]
\includegraphics[width=\linewidth]{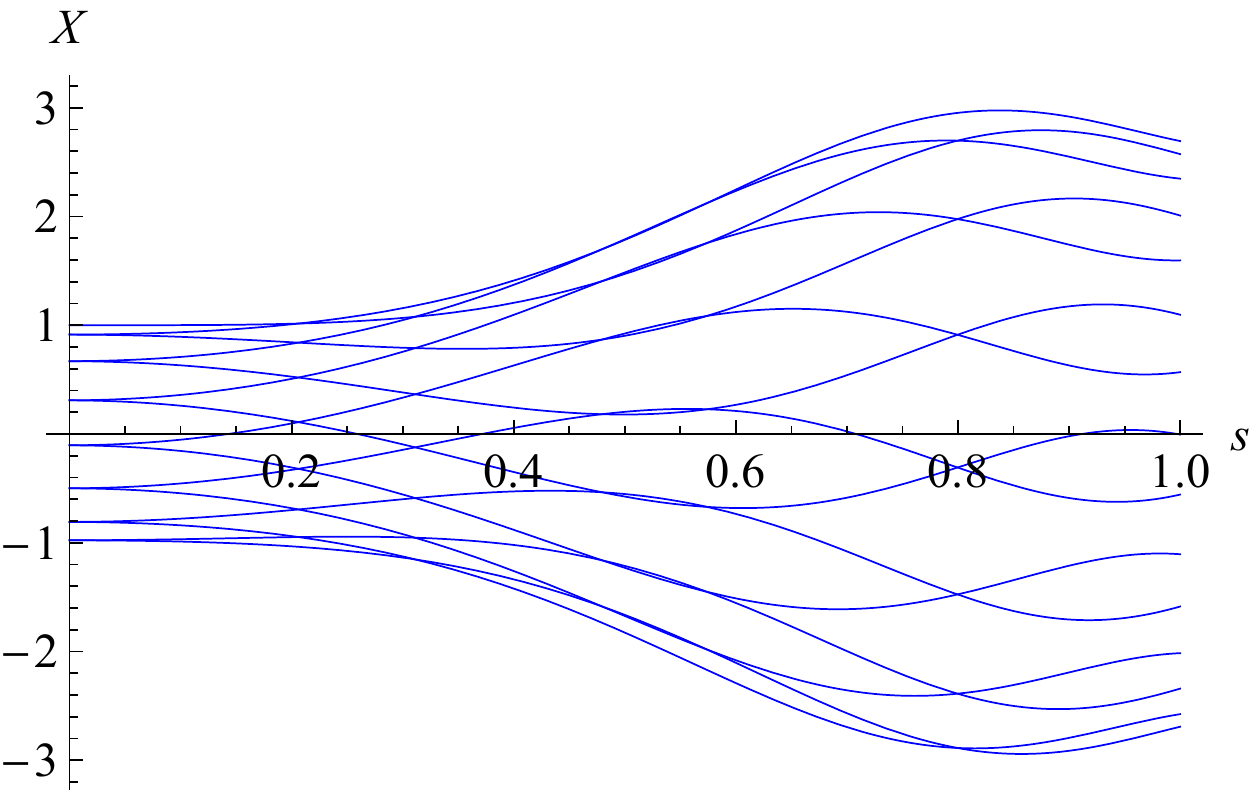}
\caption{\label{XKappa0p5Tau10}
	Same as Fig.~\ref{XKappa0Tau10}, but for the half-compensation, $\kappa=0.5$, close to the case of Ref.~\cite{An:2017FACET2}.	Amplification of the initial offset is reduced by a factor of $\sim 3$.}
\end{figure}
\begin{figure}[h!]
\includegraphics[width=\linewidth]{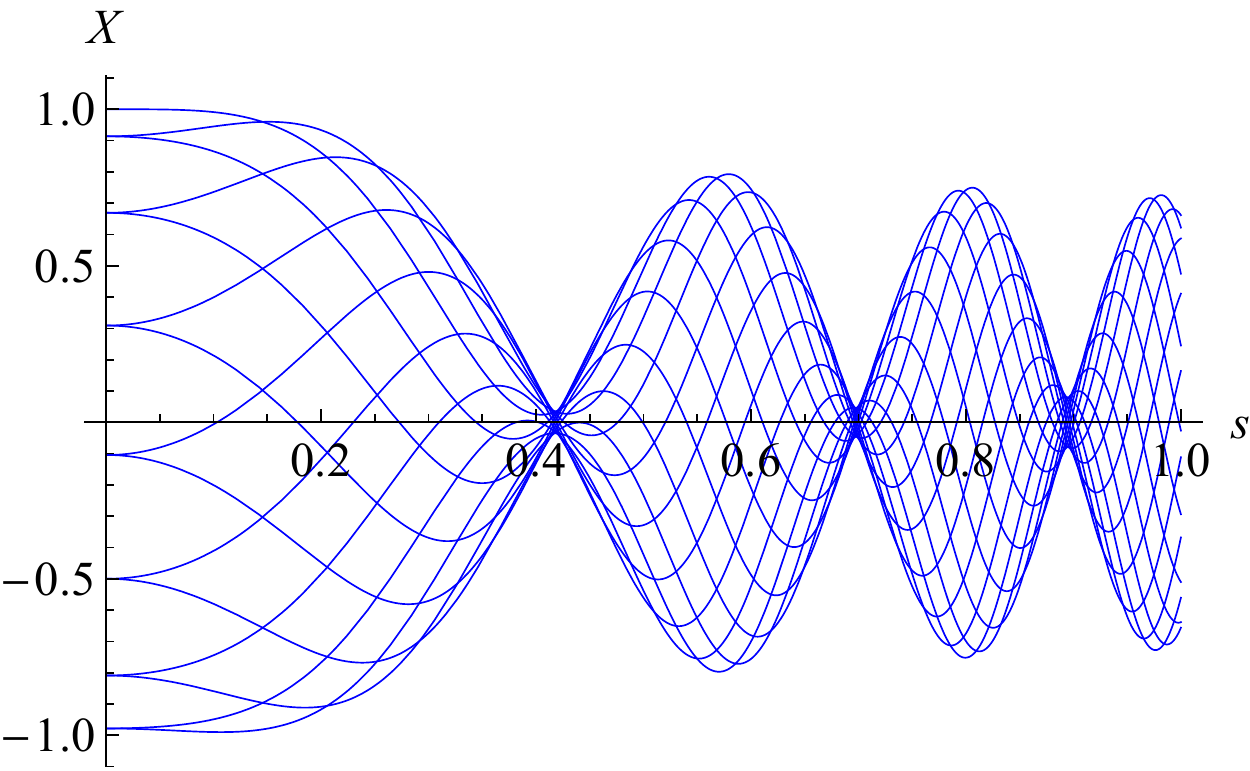}
\caption{\label{XKappa2Tau10}
Same as Figs.~\ref{XKappa0Tau10}~and~\ref{XKappa0p5Tau10}, for the double overcompensation, $\kappa=2$. 
	}
\end{figure}

Since the problem is reduced now to finding a function of just two parameters, $g_\kappa(u)$, from a linear ordinary integro-differential equation (\ref{gu}), it can be easily solved numerically for all interesting cases; Fig.~\ref{Fig:XKappaTau} presents the amplitude modulus $x(1,\tau)$ as a 3D plot for $0 \leq \kappa \leq 1.2$ and $0 \leq \tau \leq 100$. 

Stroboscopic patterns of oscillations $X(\zeta)=\Re\, x(\zeta) \exp(-i k\Delta \mu)$, with $k=1,\,2,\,3,\, ...$ and arbitrarily chosen phase $\Delta \mu$, are presented in Figs.~\ref{XKappa0Tau10}~-~\ref{XKappa2Tau10} for the specified values of the modulation parameter $\kappa$. For all the three cases $\tau=10$, being close to the simulations of Ref.~\cite{An:2017FACET2} for the planned FACET~II parameters.  

A couple of things are worth noting in relation to Figs.~\ref{Fig:XKappaTau}~--~\ref{XKappa2Tau10}.  First, the instability is considerably weakened even with a moderate frequency modulation, $\kappa \simeq 0.2-0.5$. Second, for overshooting, $\kappa >1$, the bunch is stable; its initial perturbation decoheres, certainly contributing to the emittance growth. 

Recollecting the driver-caused frequency spread, Eq.~(\ref{domegad}), a limit $\tau < \eta_t \omega_\perp/\delta \omega_d$ has to be assumed for all the computations where this spread was neglected. 

At the end, we may stress that the plasma ion mobility in the Coulomb fields of the driving and trailing bunches mitigates the beam breakup of the trailing bunch. At the same time, after certain thresholds, this mobility may lead to a dramatic emittance growth due to extremely high nonlinearity of the ion collapse~\cite{PhysRevLett.95.195002}. The obtained estimations well agree with the simulations of Ref~\cite{An:2017FACET2} and suggest new comparisons with PWA simulations and measurements.

Fermilab is operated by Fermi Research Alliance, LLC under Contract No. DE-AC02-07CH11359 with the United States Department of Energy.

\bibliography{bibfile}			

\end{document}